\documentstyle[graphicx,floats,aps,eqsecnum,preprint]{revtex} 
\tightenlines
\begin{document}

%\preprint{draft}

\newcount\eLiNe\eLiNe=\inputlineno\advance\eLiNe by -1
\title{
Gamow Shell-Model Description of Weakly Bound and Unbound Nuclear States
}
\author{N. Michel$^{1-3}$, W. Nazarewicz$^{1,2,4}$, M. P{\l}oszajczak$^{5}$,\\ and
J. Rotureau$^{5}$}
\address{$^{1}$Department of Physics and Astronomy,  University of
Tennessee\\ Knoxville, Tennessee 37996, USA}
\address{$^{2}$Physics Division, Oak Ridge National Laboratory, P.O. Box
2008\\
Oak Ridge, TN 37831, USA}
\address{$^{3}$Joint Institute for Heavy Ion Research,Oak Ridge National Laboratory 
\\Oak Ridge, Tennessee 37831}
\address{$^{4}$Institute of Theoretical Physics, Warsaw University, ul.
Ho\.{z}a 69\\PL-00681 Warszawa, Poland}
\address{$^{5}$Grand Acc\'{e}l\'{e}rateur National d'Ions Lourds (GANIL),
CEA/DSM -- CNRS/IN2P3, BP 55027, F-14076 Caen Cedex 05, France}

\maketitle

\begin{abstract}
Recently, the shell model in the complex $k$-plane (the so-called Gamow Shell
Model) has been formulated using a
complex Berggren ensemble representing bound single-particle states,
single-particle resonances, and non-resonant continuum states. 
In this framework, we shall discuss binding energies and 
  energy spectra of neutron-rich helium and lithium isotopes.
The single-particle basis used is that of the Hartree-Fock
potential generated self-consistently by the finite-range residual interaction. 
\end{abstract}


\section {Introduction}
Low-energy nuclear physics is undergoing a revival with revolutionary
progress in radioactive beam experimentation. New facilities have been
built or are under construction, and new ambitious future projects, such
as the Rare Isotope Accelerator  in the U.S.A., 
will shape research in this field 
for decades to come. From a
theoretical point of view, the major problem is to achieve a consistent picture
of weakly bound and  unbound nuclei, which requires an accurate description of
the particle continuum properties when carrying out 
 multi-configuration mixing. This is the domain of the
continuum shell model \cite{opr} and, most recently, the 
Gamow Shell Model (GSM) \cite{Mic02,Mic03} (see also \cite{betan}). 
GSM is the multiconfigurational shell model with a single-particle (s.p.)
basis given by the Berggren ensemble \cite{berggren} which contains Gamow (or
resonant) states and the complex non-resonant continuum. The resonant states
are the generalized eigenstates of the time-independent Schr\"{o}dinger
equation which are regular at the origin and satisfy purely outgoing boundary
conditions. The s.p. Berggren basis is  generated by a 
finite-depth potential, and the many-body states are obtained in
shell-model  calculations as the linear combination of Slater determinants
spanned by bound, resonant, and non-resonant s.p. basis states. Hence, both
continuum effects and correlations between nucleons are taken into 
account simultaneously. An interested reader can find all details of the
formalism in Refs. \cite{Mic02,Mic03} in which
the GSM was applied to many-neutron configurations in neutron-rich
helium and oxygen isotopes.
 In this contribution, 
we shall present the first application of the GSM formalism to the 
 $p$-shell nuclei in the model space involving both neutron and proton
s.p. Gamow states calculated from the self-consistent
Hartree-Fock (HF) method (Gamow Hartree-Fock, GHF). 
Moreover, we briefly report on the first successful application 
of the density matrix renormalization group (DMRG) 
technique\cite{white} in the context of 
realistic multiconfigurational shell model (SM).

\section {Description of the calculation}
In our previous studies \cite{Mic02,Mic03}, 
we have used the Surface Delta Interaction (SDI) and 
the s.p. basis has been generated
by a Woods-Saxon (WS) potential which was adjusted to reproduce 
the s.p. energies in $^5$He. This  potential (``$^{5}$He" parameter set
\cite{Mic03}) has the radius $R_0=2$
fm, the diffuseness $d=$0.65 fm, the strength of the central field 
$V_0=47$ MeV, and
the spin-orbit strength $V_{so}=$7.5 MeV.
The SDI interaction has several disadvantages in practical applications. 
Firstly, it has zero range, so one is forced  to introduce
an energy cutoff and, consequently, the residual interaction
depends explicitly
on the 
model space.
Moreover, one is bound to use the same WS basis for all nuclei, 
as the SDI interaction cannot practically be used to generate the  
HF  potential.
Consequently, as the chosen WS basis is not an optimal s.p. basis
(HF basis is), one  cannot easily truncate the configuration 
space when the number of valence particles increases \cite{Mic03}.  
So, we have decided to introduce  a new two-body residual interaction, the 
Surface Gaussian Interaction (SGI): 
\begin{equation}
V_{J,T}(\vec{r_1},\vec{r_2}) = V_0(J,T) \cdot \exp \left[ - \left( \frac{ \vec{r_1}-\vec{r_2} }{\mu} \right) ^2 \right]
\cdot \delta{ \left( |\vec{r_1}| +  |\vec{r_2}| - 2 \cdot R_0 \right) },
\label{eq1}
\end{equation}   
which is used together with the WS potential
with the ``$^{5}$He" parameter set.

The SGI interaction
is a compromise between  the SDI and the Gaussian 
interaction. The parameter
$R_0$ in Eq.~(\ref{eq1}) is the radius of the WS potential, and 
$V_0(J,T)$ is the coupling constant which explicitly depends
on the total angular momentum $J$ and and the total isospin 
$T$ of the pair of nucleons.
A principal  advantage of the SGI is that it 
is finite-range, so no  energy cutoff is needed.
Moreover, the surface delta term in (\ref{eq1}) simplifies 
the calculation of two-body matrix elements,
because the radial integrals become one-dimensional and they extend from $r$=0
to $r$=2$R_0$.
 (In the Gaussian case, such as the Gogny force,  they are two-dimensional 
and have to be extended to infinity.) 
Consequently, an adjustment of the Hamiltonian parameters
becomes feasible. Finally, the resulting spherical HF 
potential 
is continuous and can be calculated very accurately. 
This allows  one to use the optimal 
spherical HF potential for the 
generation of the Berggren basis for each  nucleus studied; hence 
a more efficient
truncation in the space of configurations with a different 
number of particles in
the non-resonant continuum. In the present study, it turned out to be
sufficient to consider at most two particles in the GHF continuum. 
This restriction on the number of particles in the non-resonant continuum
allowed us to extend the studies up to unbound 
$^{10}$He and  the halo nucleus $^{11}$Li.

\subsection {Choice of the valence space}

In our He and Li calculations,
the valence space for protons and neutrons consists of 
$0p_{3/2}$ and $0p_{1/2}$ spherical  GHF resonant states, calculated for  each nucleus,
 and the 
$\{ ip_{3/2}\} $ and $\{ ip_{1/2}\} $ $(i=1,\cdots ,M)$ complex continua
generated by the same potential. These continua extend from $Re[k]$=0 to 
$Re[k]$=8\,fm$^{-1}$ and they are  discretized with 14 points 
(i.e., $M=14$). Altogether, we have 
15 $p_{3/2}$ and 15 $p_{1/2}$  GHF states (shells) 
in the GSM calculation. The imaginary parts of $k$-values of the discretized 
continua are chosen to minimize the error
made in calculating the imaginary parts of energies of the many-body states.
Other continua, such as  $s_{1/2}$, $d_{5/2}$, $\cdots$ are neglected, 
as they can be chosen to be real and would only induce
a renormalization of the two-body interaction. We have checked  
\cite{Mic02,Mic03} that their influence on the binding energy of light 
helium isotopes is negligible. On the other hand, the $1s_{1/2}$ anti-bound 
neutron s.p. state is important in the heaviest Li isotopes ($^{10}$Li, $^{11}$Li)
and plays a significant role in explaining the halo ground-state
(g.s.) configuration of 
$^{11}$Li \cite{thompson,Bet03}. At present, however, solving a GSM problem 
for $^{11}$Li  in $[0p_{3/2},\{ ip_{3/2}\}; 
0p_{1/2},\{ ip_{1/2}\}; 1s_{1/2},\{ is_{1/2}\}; (i=1,\cdots ,M)]$ 
GHF space
is not possible within a reasonable computing time. This
task will be, however, possible in the near  future 
by using a new generation GSM code which employs the
DMRG methods 
to include the non-resonant continuum configurations
contribution in the many-body wave function \cite{jimmy1,jimmy}. 

Having defined a discretized GHF  basis, one constructs
the Slater determinants from all s.p. basis states 
(bound, resonant, and non-resonant), 
keeping only those with at most two particles in the non-resonant continuum.
 Indeed, as the two-body Hamiltonian is
diagonalized in its optimal GHF basis, the weight of configurations involving
 more 
than two particles in the continuum is usually quite small,
and they are neglected in the following.

\subsection {The Helium chain}\label{HeHe}

Within the chain of helium isotopes, which are described assuming
an inert $^{4}$He core, there are  only $T$=1 two-body matrix elements. 
Consequently, only ($J$=0, $T$=1) and ($J$=2, $T$=1) couplings
 come into play. We have adjusted $V_0(J=0,T=1)$ 
to reproduce the experimental g.s. energy of $^6$He 
relative to the g.s. of $^{4}$He, 
whereas $V_0(J=2,T=1)$ has been fitted to
all g.s. energies from $^7$He to $^{10}$He. Indeed, these latter states 
are mainly sensitive to $V_0(J=2,T=1)$ whereas, for obvious geometrical 
reasons,  the $J=2,T=1$ coupling is absent in the g.s. of $^6$He.
The adopted  values are: $V_0(J=0,T=1)$ = --403 MeV fm$^3$ and 
$V_0(J=2,T=1)$ = --315 MeV fm$^3$. 

The calculated spectrum of He isotopes is shown in Fig.~\ref{He_spectrum}. 
The experimental g.s. binding energies relative to the 
$^{4}$He core are  reproduced fairly well.
For instance,
 the g.s. of $^{6}$He and $^{8}$He are 
bound whereas  $^{5}$He and $^{7}$He are unbound. Moreover,
the so-called `helium anomaly' (see, e.g., Ref. \cite{oglo}), 
[i.e.,  the higher one-
and two-neutron emission thresholds in $^{8}$He than in $^{6}$He], 
is well reproduced. The qualitative features seen 
in the experimental spectra are also satisfactorily reproduced by the 
GSM with the SGI
Hamiltonian. The width of certain resonance states is too
large mainly because these states are calculated  at too high an excitation 
energy above the one- and two-nucleon emission threshold.

Looking at the configuration mixing in the g.s. of helium isotopes
(see Tables \ref{tables_GS_He6}-\ref{tables_GS_He10}), 
one can see that the coupling to the non-resonant continuum
is most important in $^6$He and $^7$He. This finding 
is consistent with the Borromean nature of $^6$He,
which is bound only because of correlation effects 
(pairing scattering to the
continuum). The g.s. of $^7$He is not a simple  s.p. resonance,
 and this
is reflected in GSM wave functions by large amplitudes of
configurations involving particle(s) in the non-resonant continuum. 
This in turn implies that the g.s. width of $^{7}$He is a
result of the complicated mixture of different configurations 
with three valence neutrons in various resonant and non-resonant shells.

The coupling to the continuum is somewhat less important in
the heavier isotopes  ($^8$He, 
$^9$He, and $^{10}$He). This is partially related to the fact that 
these nuclei have  a 
closed $0p_{3/2}$ shell; hence for the bound g.s. of $^8$He,
 the calculated 
spherical HF potential is the {\em exact} HF potential (no deformation
effects are present). Consequently, the
contributions of the  1p-1h excitations to the g.s. wave function 
vanish.
For unbound states, 
there are small 1p-1h components due to the 
neglect of the imaginary part in the approximate HF
potential used in the actual calculation.
 Secondly, in the heavier He isotopes, the GHF $0p_{3/2}$  shell is calculated 
to be  bound, which
 further  diminishes the importance of the coupling to 
 the non-resonant continuum.
Nevertheless, this coupling is by no means negligible,
as the configurations involving states of
the non-resonant continuum  still represent about 10\%
of the wave function and play a significant role in 
generating  binding  
for all these nuclei.

Comparing the SGI results with those of Ref. \cite{Mic03}, one can 
see that the main conclusions  for $^6$He and $^7$He still hold. This
is because the HF potential for these nuclei is close to 
the WS potential which was  previously used.
 However, once the $0p_{3/2}$ shell gets closed, 
the HF field becomes very different
from the WS potential. As a consequence, the large 
configuration mixing obtained in Ref.~\cite{Mic03} in the WS basis
for $^8$He  is significantly reduced when the 
GHF  basis is used. In other words, a large part of the configuration mixing 
in the WS basis is not due to the presence of genuine two-particle 
correlations involving continuum states,
but due to the 1p-1h couplings now incorporated in the GHF basis.
                                                           
In order to assess  the importance of two-particle correlations involving 
continuum states,
we have calculated the g.s. of $^6$He, $^7$He, and $^8$He, 
but in the model space of  1p-1h excitations
only. In such truncated calculations,  the real part of the energy reads:
0.75 MeV for  $^{5}$He,  0.58 MeV 
for $^6$He, 1.27 MeV 
for  $^7$He, and  --1.40 MeV 
for $^8$He.
Based on these numbers, one is tempted to conclude that 
the anomalous increase of one- and two-neutron separation
energies in $^{8}$He, as compared to $^{6}$He, is a genuine mean-field effect
 caused by the 
$J=2,T=1$ coupling. 
The coupling to the non-resonant continuum further enhances  
the helium  anomaly. 

The drip-line nucleus $^8$He is two-neutron 
 bound already in the HF approximation, so the 
coupling to the non-resonant continuum provides only  
additional binding. On the other hand, $^6$He 
 is unbound in  GHF, so the continuum coupling is solely 
responsible for binding. Therefore, one may conclude that 
the Borromean features in the  $^{4-8}$He chain
are caused mainly by the continuum couplings, 
whereas the helium anomaly is mainly due to the  change from a 
($J=0,T=1$)-dominated  mean field to  a   mean field dominated by the 
($J=2,T=1$) coupling.

\subsection {The Lithium chain}     
As a first example of GSM calculations in the space of proton {\it and} 
neutron states, 
we have chosen to investigate  the Li chain. The
continuum effects are very important in these nuclei, 
both in their  ground  states and in excited states. 
The nucleus $^{11}$Li is also a 
well-known example of a two-neutron halo. In our $p$-space(s) calculation,
 we  consider the  one-body Coulomb potential of the $^{4}$He core, 
which is given by a uniformly charged sphere
having  the radius of the WS potential. It turns out that the inclusion of
the one-body Coulomb potential  modifies the GHF basis in
lithium isotopes as compared to the helium isotopes, an effect which is
usually neglected in the standard SM calculations.

Recent studies of the binding energy systematics in the
$sd$-shell nuclei using the
Shell Model Embedded in the Continuum (SMEC) have reported
a significant reduction of the neutron-proton $T$=0
 interaction with respect
to the neutron-neutron $T$=1 interaction \cite{luo,mazury} in the nuclei
close to the neutron drip line. In SMEC  \cite{smec}, 
this reduction is associated with  a  decrease in the
one-neutron emission threshold when approaching the neutron drip line, 
i.e.,  it is a genuine continuum coupling effect. The
detailed studies in fluorine isotopes have shown that the  reduction of 
the $T$=0
neutron-proton interaction {\em cannot} be corrected by 
any adjustment of the monopole components  of the effective
Hamiltonian. To account for  this effect  in the standard 
SM, one would need to introduce a $N$-dependence of the $T$=0 
monopole terms.
Interestingly, it has recently been suggested \cite{zuker}
that a linear reduction of $T$=0 two-body 
monopole terms  is expected if one incorporates  three-body
interactions into  the two-body framework of a standard SM. 

Our  GSM studies of lithium isotopes indicate that the reduction of
$T$=0 neutron-proton interaction with increasing neutron number is
essential. For example, if one uses the $V_0(J,T=0)$ strength adjusted 
to $^{6}$Li to calculate $^{7}$Li, the g.s. of $^{7}$Li becomes overbound by 13
MeV, and the situation becomes even worse for heavier Li isotopes. To reduce this
disastrous tendency, in the first approximation we have used a linear dependence
of $T$=0 couplings on the number of valence neutrons $n$:
\begin{eqnarray}
\label{inter}
V_0(J=1,T=0)=\alpha_{10}\left[1-\beta_{10}(n-1)\right],  \nonumber \\ \\
V_0(J=3,T=0)=\alpha_{30}\left[1-\beta_{30}(n-1)\right], \nonumber
\end{eqnarray}
with $\alpha_{10}=-600$ MeV fm$^3$, $\beta_{10}=-50$ MeV fm$^3$,
$\alpha_{30}=-625$ MeV fm$^3$, and $\beta_{30}=-100$ MeV fm$^3$. 
This linear dependence is probably oversimplified, as shown in
Refs.~\cite{luo,mazury} where the proton-neutron $T$=0 interaction first
decreases fast with increasing neutron number and then saturates for weakly
bound systems near the neutron drip line. For the $T$=1
interaction, we have taken the parameters $V_0(J=0,T=1)$ and 
$V_0(J=2,T=1)$ determined for the He chain 
(see Fig. \ref{He_spectrum} and the preceding discussion). 

\begin{figure}[htbp]
\begin{center}
\includegraphics[height=11.5cm]{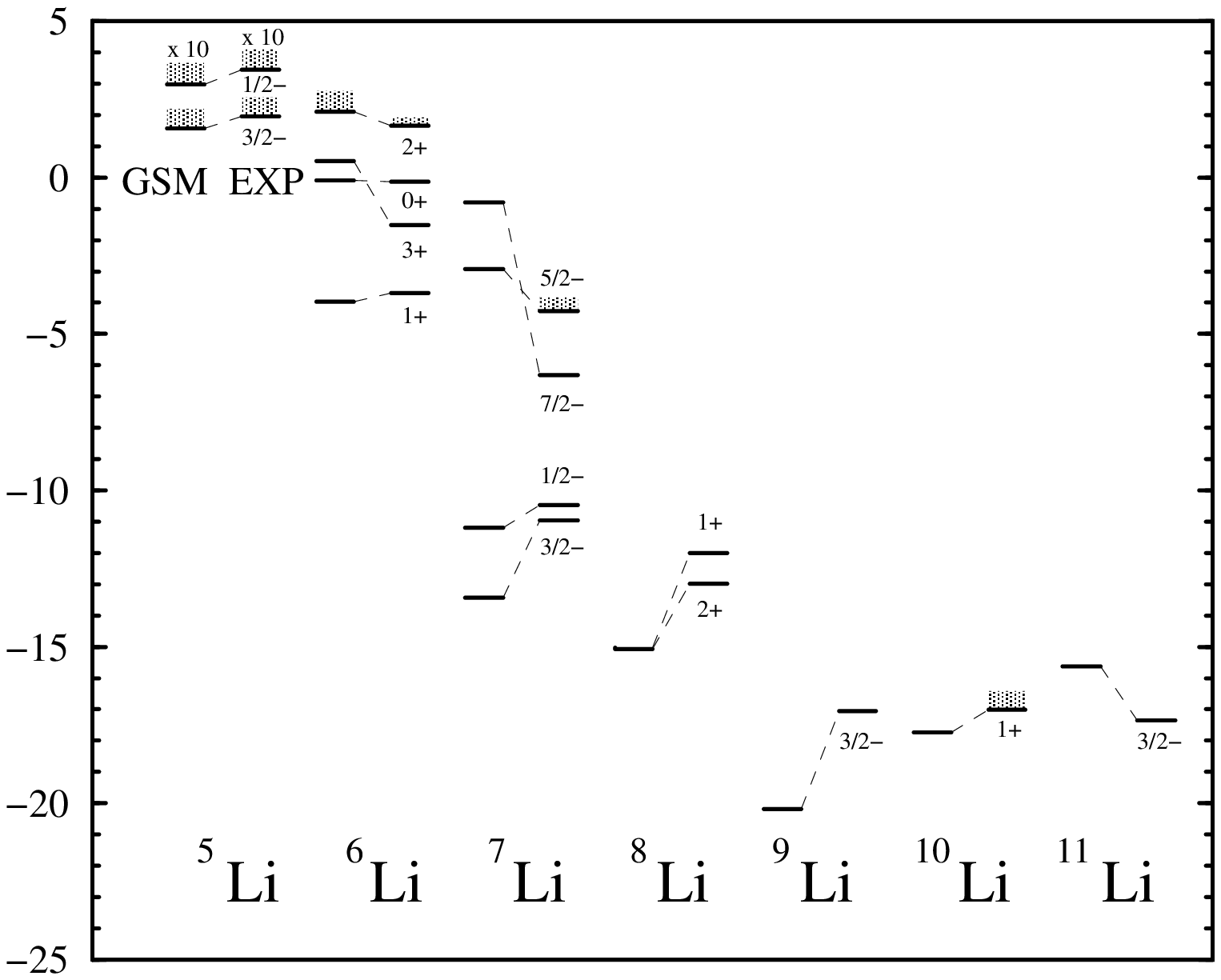}
\end{center}
\caption{GSM spectra of lithium isotopes obtained with the SGI Hamiltonian.
Experimental data are taken from 
Refs.~\protect\cite{ensdf,audi_wapstra,bohlen_Li10}.}
%WN
%WN Experimental References
%WN
\label{Li_spectrum}
\end{figure}

The results of our GSM calculations
for the neutron-rich Li isotopes are shown in  Fig.~\ref{Li_spectrum}. 
One obtains a reasonable description of the g.s. energies of lithium isotopes 
relative to the g.s. energy of  $^{4}$He, even though the agreement with
the data is somewhat worse than in the 
He chain. Clearly, the particle-number dependence of the $T$=0 matrix 
elements has to be
further investigated. The absence of an antibound $s_{1/2}$ state 
in the Berggren basis is most likely
 responsible for large deviations with the data seen in $^{10}$Li and
$^{11}$Li. 

\section{Future perspectives: the Density Matrix Renormalization Group
techniques for solving the Gamow Shell Model}
As outlined above, in the GSM one uses a Berggren basis which consists of
discrete states (bound and resonant states, $l=0$ anti-bound state for 
neutrons)  and  the 
discretized non-resonant 
continuum. Consequently, the dimension of the (non-hermitian) Hamiltonian
matrix in GSM grows extremely fast with increasing the size of  the Hilbert space.
This `explosive' growth of the
dimension is much more severe than 
in the standard SM for which the dimensionality problem concerns only the 
`pole' space of GSM. The future perspectives of GSM applications are ultimately
related to the progress in developing  new methods of truncating
 huge SM  spaces. One  promising approach is the DMRG method. In nuclear structure,
  this  method has been successfully applied in  schematic
 Hamiltonians \cite{duke}, but no fully convincing results have  so far
 been reported in the context of the  realistic SM.

Let us begin by summarizing  the  basic elements of  DMRG  
\cite{white} (see also Ref. \cite{duke} for a pedagogical discussion relevant
to  nuclear physics applications). 
The main idea is to consider `step by step' different s.p. 
shells in the configuration
space and retain  only the  $N_{opt}$ ``best states'' 
dictated by
the one-body density matrix. The convergence of the DMRG method is then
studied with respect to $N_{opt}$. 

The configuration space is divided into two subspaces
denoted by ${\cal H}$ and ${\cal P}$. Generally speaking, ${\cal H}$ contains 
the lowest s.p. shells. 
In the first step, 
 one calculates and stores all the possible matrix elements of suboperators
of the Hamiltonian in the ${\cal H}$ space:
$$a^{\dagger }, (a^{\dagger }\, \widetilde{a})^{j}, 
(a^{\dagger }a^{\dagger })^{j}, 
((a^{\dagger }a^{\dagger })^{j} \widetilde{a})^{J}), 
((a^{\dagger }a^{\dagger })^{j} (\widetilde{a}\widetilde{a})^{j})). $$
One also constructs in ${\cal H}$ all states $|h\rangle$ 
with $0,1,2,\cdots $ particles
coupled to all possible $J$-values.
Then, one considers the first s.p. shell in ${\cal P}$, 
calculates matrix elements of all suboperators in
this shell, and constructs in ${\cal P}$ all states $|p\rangle$ with 
$0,1,2,\cdots $ particles coupled to all possible $J$-values.
 In the following, one adds  `one by one'  additional  s.p. shells in 
${\cal P}$:  one calculates all matrix  elements of
suboperators in the added shell and all states with $0,1,2,\cdots $ 
particles in the added shell and in the shell previously considered in ${\cal
P}$. New states  are successively added in ${\cal P}$ until the number of states 
$|p\rangle$ with $0,1,2,\cdots $ particles is larger than $N_{opt}$. 
Then one diagonalizes the Hamiltonian in the space $|hp\rangle^{J}$ made of
vectors in ${\cal H}$ and ${\cal P}$. Obviously, the number of particles 
in such states is equal to the total number of valence particles in 
the system, and $J$ is equal to the angular momentum of the 
state of interest. From the eigenstates : 
\begin{equation}
\Psi =\sum c_{hp}|hp\rangle^{J},
\end{equation}
one calculates the one-body density matrix:
\begin{eqnarray}
\rho _{pp'}=\sum _{h} c_{hp}c_{hp'}.
\end{eqnarray}
The density matrix is diagonalized and only $N_{opt}$ 
eigenstates having the largest eigenvalues are retained. 
(In the non-hermitian
GSM problem, eigenvalues of the density matrix are complex and the eigenstates
are selected according to the largest {\em absolute} value of the density 
eigenvalues.) One then recalculates all the matrix elements of suboperators 
for the optimized states; they are linear combinations of  
previously calculated  matrix
elements. Then, one adds the next shell in ${\cal P}$ and, again, only 
the $N_{opt}$ 
states selected according to the above prescription are kept.
This procedure is continued until the last shell in ${\cal P}$ is reached,
providing a `first guess' for the many-body wave function.

At this stage, a ``sweeping phase" begins in which the 
iterative process is reversed. First, 
one  considers the last shell in ${\cal P}$. At this point,   one
constructs states with $0,1,2,\cdots $ particles and then the 
process  continues, step by step, 
until the number of vectors  becomes larger than $N_{opt}$. 
When  the $i^{th}$ shell
in ${\cal P}$ is reached, the Hamiltonian is diagonalized 
in the set of vectors 
$|h, p_{prev}, p\rangle^{J}$, where $h$ is a state that belongs to ${\cal H}$, 
$p_{prev}$ is a previously optimized state 
($i$--1 first shells in ${\cal P}$), and $p$ is a state 
that concerns all the shells between  the $i^{th}$ one and
 the last one in ${\cal P}$. 
The density
matrix is then diagonalized and  the $N_{opt}$  $p$-states are kept.
The procedure continues by  adding the 
$(i-1)^{st}$ shell in ${\cal P}$, etc., until  the first state is reached. 
Then the procedure is reversed:  the
first shell is added, then the second, the third, etc. 
The succession of sweeps is successful if
the energy gradually converges after every sweep.

As a  first application of DMRG, we  calculated
the  g.s. $0_1^+$ and the excited $0_2^+$ 
state in  $^{6}$He in the  same  configuration space as in Sec.~\ref{HeHe}.
Initially, we applied the DMRG procedure
without sweeping, i.e.,  in its  `infinite algorithm' version
\cite{white}. In this calculation, 
we considered 14 $p_{1/2}$ shells and 14 $p_{3/2}$ shells in the non-resonant 
continuum. The  method   succeeded in
reproducing the  ``exact" g.s. energy of  $^{6}$He  when taking  16 shells in
${\cal H}$ (8 $p_{1/2}$ shells and 8 $p_{3/2}$ shells) and keeping
 6 vectors in ${\cal P}$ at each iteration.
One should mention that the states in ${\cal H}$ have not been 
optimized during the iterative procedure. This example  demonstrates that
the DMRG in the  infinite-system variant  can be generalized for genuinely
non-hermitian problems. On the other hand, the resulting gain 
 in reducing the dimensionality of the GSM   is not particularly impressive. 

The gain factor is
radically improved when  the  finite-system algorithm 
(sweeping) is applied. In this case,
for 20 $p_{1/2}$ shells and 20 $p_{3/2}$ shells in the non-resonant
continuum,  an excellent 
convergence has been reached  for 
both states of $^6$He by taking  $0p_{1/2}$ and $0p_{3/2}$
Gamow resonances   in ${\cal H}$   and keeping in ${\cal P}$ only 
six  vectors after adding each shell. In the considered example,
the total dimension of the GSM
Hamiltonian  is 462 and the rank of the
biggest matrix to be
diagonalized in GSM+DMRG is 32. The gain factor is expected to be even more
 impressive for a larger number of valence particles.

\section{Conclusions}
The Gamow Shell Model, which has been 
introduced only very recently \cite{Mic02,Mic03}, has proven to be a reliable tool
for the microscopic description of weakly bound and unbound nuclear states. In He 
isotopes, GSM with either SDI or SGI interactions was able to describe 
fairly well
the many-body properties, in particular the Borromean features in
the chain $^{4-8}$He. Using the finite-range SGI interaction made it 
possible to perform GHF calculations, thus  designing the optimal
Berggren basis for each nucleus. In this way, we were able to
disentangle the correlations due to the continuum coupling from the 
particle-hole excitations
essential for building the mean field. We have shown that the Borromean
features of the helium isotopes are the results of the correlations involving
the non-resonant continuum, whereas the helium anomaly is a
 mean field effect due to the transition from a ($J=0,T=1$)-dominated 
 mean field
 in $^{6}$He to the ($J=2,T=1$)  GHF field in $^{8}$He.
 
In Li isotopes, $T$=0 matrix elements of the two-body interaction could be
studied for bound and resonant many-body states. It was found that the $T$=0
interaction contains a pronounced density (particle-number)
dependence which originates from the
coupling to the continuum and leads to an effective renormalization of the
neutron-proton coupling. This effect cannot be absorbed 
by the modification of $T$=0 monopole terms in the standard SM
framework. The effective 
renormalization of ($J=1,T=0$) and ($J=3,T=0$) couplings, 
which has been found in
the present GSM studies, has to be further investigated, as it is also related
to the question about the importance of 3-body  correlations
and density dependence. 

The successful application of GSM  to heavier nuclei is ultimately related to
the progress in  optimization of the GSM basis. The promising 
development, discussed in this paper, is the adaptation of
the DMRG method \cite{white} to the genuinely
non-hermitian SM problem in the complex-$k$ plane. The first  
applications using the $j$-scheme GSM are very promising; they
demonstrate that we may well be at the edge of solving the GSM in today's
inaccessible model spaces. 

\vskip 0.5truecm

This work was supported in part by the U.S.\ Department of Energy under
Contract Nos. \ DE-FG02-96ER40963 (University of Tennessee) and
DE-AC05-00OR22725 with UT-Battelle, LLC (Oak Ridge National Laboratory),
DE-FG05-87ER40361 (Joint Institute for Heavy
Ion Research),
and by the Polish Committee for Scientific Research (KBN).

\newpage

\begin{table}[hbtp]
\caption{Squared amplitudes of configurations in the ground 
state of $^{6}$He.
The sum of squared amplitudes of Slater determinants with 
$n$ particles in the continuum is denoted by $L_{+}^{(n)}$.}
\label{tables_GS_He6}
\vskip 0.5truecm     
\begin{center}
\begin{tabular}{|c|c|} 
Configuration & $c^2$ \\ \hline 
$0p^{2}_{3/2}$ & {0.656--i0.566} \\ 
$0p^{2}_{1/2}$ & {6.06$\cdot 10^{-3}$--i0.0516} \\ 
$L_{+}^{(1)}$   & {0.363+i0.509} \\ \hline 
$L_{+}^{(2)}$   & {--0.0245+i0.108} 
\end{tabular}
\end{center}
\end{table}

\begin{table}[hbtp]
\caption{Similar as in Table~\protect\ref{tables_GS_He6} except for the
$3/2^-_1$ g.s.
of $^7$He.}
\vskip 0.5truecm     
\label{tables_GS_He7}
\begin{center}
\begin{tabular}{|c|c|} 
Configuration & $c^2$ \\ \hline 
$0p^{3}_{3/2}$ & {0.331--i0.0973} \\ 
$0p^{1}_{3/2}$ $0p^{2}_{1/2}$ & {0.0111--i0.0406} \\ 
$0p^{2}_{3/2}$ $0p^{1}_{1/2}$ & {5.380$\cdot 10^{-4}$--i6.950$\cdot 10^{-4}$} \\ 
$L_{+}^{(1)}$   & {0.507+i0.0714} \\ 
$L_{+}^{(2)}$   & {0.150+i0.0672} 
\end{tabular}
\end{center}
\end{table}

\begin{table}[hbtp]
\caption{Similar as in Table~\protect\ref{tables_GS_He6} except for the g.s.
of $^8$He.}
\vskip 0.5truecm     
\label{tables_GS_He8}
\begin{center}
\begin{tabular}{|c|c|} 
Configuration & $c^2$ \\ \hline 
$0p^{4}_{3/2}$ & {0.889--i7.826$\cdot 10^{-3}$} \\ 
$0p^{2}_{3/2}$ $0p^{2}_{1/2}$ & {0.0316--i0.0529} \\ 
$L_{+}^{(1)}$   & {0.0613+i0.0226} \\ 
$L_{+}^{(2)}$   & {0.0184+i0.0225} 
\end{tabular}
\end{center}
\end{table}

\begin{table}[hbtp]
\caption{Similar as in Table~\protect\ref{tables_GS_He6} except for the
$1/2^-_1$ g.s.
of $^9$He.}
\vskip 0.5truecm     
\label{tables_GS_He9}
\begin{center}
\begin{tabular}{|c|c|} 
Configuration & $c^2$ \\ \hline 
$0p^{4}_{3/2}$ $0p^{1}_{1/2}$ & {0.937--i0.0209} \\ 
$L_{+}^{(1)}$   & {0.0473--i1.779$\cdot 10^{-3}$} \\ 
$L_{+}^{(2)}$   & {0.0157+i0.0227} 
\end{tabular}
\end{center}
\end{table}

\begin{table}[hbtp]
\caption{Similar as in Table~\protect\ref{tables_GS_He6} except for the g.s.
of $^{10}$He.}
\vskip 0.5truecm     
\label{tables_GS_He10}
\begin{center}
\begin{tabular}{|c|c|} 
Configuration & $c^2$ \\ \hline 
$0p^{4}_{3/2}$ $0p^{1}_{1/2}$ & {0.965--i0.0206} \\
$L_{+}^{(1)}$   & {-5.640$\cdot 10^{-3}$+i8.392$\cdot 10^{-3}$} \\ 
$L_{+}^{(2)}$   & {0.0409+i0.0122} 
\end{tabular}
\end{center}
\end{table}
             
\newpage

\begin{figure}[htbp]
\begin{center}
\includegraphics[height=11.5cm]{SGI_spectrum.eps}
\end{center}
\caption{GSM spectra of helium isotopes obtained with the SGI Hamiltonian.
Experimental data are taken from 
Refs.~\protect\cite{ensdf,audi_wapstra,meister,bohlen_he10}.}
\label{He_spectrum}
\end{figure}
  
\newpage

\end{document}